\documentclass[10pt]{article}
\usepackage{graphicx}
\usepackage{amsmath}
\usepackage{amssymb}
\usepackage{balance}
\usepackage{bm}
\usepackage{caption2}
\usepackage{indentfirst}

\begin{document}
\bibliographystyle{prsty}
\begin{center}
{\large {\bf \sc{ Analysis of the $D\bar{D}^*K$ system with QCD sum rules }}} \\[2mm]
Zun-Yan Di$^{1,2}$, Zhi-Gang Wang$^{1}$\footnote{E-mail: zgwang@aliyun.com. }   \\
$^{1}$ Department of Physics, North China Electric Power University, Baoding 071003, P. R. China \\
$^{2}$ School of Nuclear Science and Engineering, North China Electric Power University, Beijing 102206, P. R. China
\end{center}

\begin{abstract}
In this article, we construct the color singlet-singlet-singlet interpolating current with $I\left(J^P\right)=\frac{3}{2}\left(1^-\right)$
to study the $D\bar{D}^*K$ system through QCD sum rules approach.
In calculations, we consider the contributions of the vacuum condensates up to dimension-16 and employ the formula $\mu=\sqrt{M_{X/Y/Z}^{2}-\left(2{\mathbb{M}}_{c}\right)^{2}}$ to choose the optimal energy scale of the QCD spectral density.
The numerical result $M_Z=4.71_{-0.11}^{+0.19}\,\rm{GeV}$ indicates that there exists a resonance state $Z$ lying
above the $D\bar{D}^*K$ threshold to saturate the QCD sum rules.
This resonance state $Z$ may be found by focusing on the channel $J/\psi \pi K$
of the decay $B\longrightarrow J/\psi \pi \pi K$ in the future.
\end{abstract}

 PACS number: 12.39.Mk, 12.38.Lg

Key words: Resonance state,  QCD sum rules

\section{Introduction}
Since the observation of the $X(3872)$ by the Belle collaboration in 2003 \cite{X3872}, more and more exotic hadrons have been observed and confirmed experimentally,
such as the charmonium-like $X$, $Y$, $Z$ states, hidden-charm pentaquarks, etc \cite{XYZ,RMP90,EPJC1534}.
Those exotic hadron states, which cannot be  interpreted as the quark-antiquark mesons or three-quark baryons in the naive quark model \cite{Quark-Model},
are good candidates of the multi-quark states \cite{PR639,PLB753}. The  multi-quark states  are  color-neutral objects because of the color confinement, and provide an important platform
to explore the low energy behaviors of QCD, as no free particles  carrying  net color charges have ever been experimentally observed. Compared to the conventional hadrons, the dynamics of the multi-quark states is poorly understood and calls for more works.

Some  exotic hadrons can be understood as  hadronic molecular states \cite{GuoRMP}, which are analogous to the deuteron as a loosely bound state of the proton and neutron.
The most impressive example is the original exotic state, the $X(3872)$, which has been studied as the $D\bar{D}^*$ molecular state by many theoretical groups \cite{PRC69,group}. Another impressive example is the $P_c(4380)$ and $P_c(4450)$ pentaquark states observed by the LHCb collaboration in 2015,
which are good candidates for the $\bar{D}\Sigma_c^{*}$, $\bar{D}^{*}\Sigma_c$,  $\bar{D}^{*}\Sigma_c^*$ molecular states \cite{GuoRMP}.
In additional to the meson-meson type and meson-baryon type molecular state, there maybe also exist meson-meson-meson type molecular states,  in other words,
there maybe exist three-meson hadronic molecules.

In Refs.\cite{KDD-1,KDD-2}, the authors explore  the possible  existence of  three-meson system $D\bar{D}^*K$  molecule according  to the attractive interactions of the two-body subsystems $DK$, $\bar{D}K$, $D^*K$, $\bar{D}^*K$ and $D\bar{D}^*$ with the Born-Oppenheimer approximation and the fixed center approximation, respectively.
In this article, we study the $D\bar{D}^*K$ system with QCD sum rules.

The QCD sum rules method is a powerful tool in studying the exotic hadrons \cite{QCD-approach,PRD83,CTP54,CTP191}, and has given many successful descriptions, for example, the mass and width of the $Z_c(3900)$ have been successfully reproduced as an axialvector tetraquark state \cite{axialvector, WangZhang-Solid}.
In QCD sum rules,
we expand the time-ordered currents
into a series of quark and gluon condensates via the operator product expansion method.
These quark and gluon condensates
parameterize the non-perturbative properties of the QCD vacuum.
According to the quark-hadron duality,
the copious information about the hadronic parameters can be obtained
on the phenomenological side \cite{hadron-side-1,hadron-side-2}.

In this article, the color singlet-singlet-singlet interpolating current with $I\left(J^P\right)=\frac{3}{2}\left(1^-\right)$
is constructed to study the $D\bar{D}^*K$ system.
In calculations, the contributions of the vacuum condensates are considered up to dimension-16 in the operator product expansion
and the energy-scale formula $\mu=\sqrt{M_{X/Y/Z}^{2}-\left(2{\mathbb{M}}_{c}\right)^{2}}$ is used to seek the ideal energy scale of the QCD spectral density.

The rest of this article is arranged as follows: in Sect.2, we derive the QCD sum rules for the mass and pole residue of the $D\bar{D}^*K$ state;
 in Sect.3, we present the numerical results and discussions; Sect.4 is reserved for our conclusion.

\section{QCD sum rules for the $D\bar{D}^*K$ state}
In QCD sum rules, we consider the two-point correlation function,
\begin{eqnarray}\label{correlation-function}
\Pi_{\mu\nu}\left(p\right)&=&i\int d^{4} x e^{ip\cdot x}
\langle0|T\left\{J_\mu(x) J_\nu^{\dag}(0)\right\}|0\rangle\, ,
\end{eqnarray}
where
\begin{eqnarray}\label{current}
J_\mu(x)&=&\bar{u}^m(x)i\gamma_5 c^m(x)\bar{c}^n(x)\gamma_\mu d^n(x)\bar{u}^k(x)i\gamma_5 s^k(x)\,,
\end{eqnarray}
the $m$, $n$, $k$ are color indexes. The color singlet-singlet-singlet current operator $J_\mu(x)$
has the same quantum numbers $I\left(J^P\right)=\frac{3}{2}\left(1^-\right)$ as the $D\bar{D}^*K$ system.

On the phenomenological side,
a complete set of intermediate hadronic states,
which has the same quantum numbers as the current operator $J_\mu(x)$,
is inserted into the correlation function $\Pi_{\mu\nu}\left(p\right)$
to obtain the hadronic representation \cite{hadron-side-1,hadron-side-2}.
We isolate the ground state contribution $Z$ from the pole term,
and get the result:
\begin{eqnarray}\label{correlation-function-1}
\Pi_{\mu \nu}\left(p\right)
&=&\frac{\lambda_Z^2}{M_Z^2-p^2}\left(-g_{\mu \nu}+\frac{p_\mu p_\nu}{p^2}\right)+\cdots \nonumber\\
&=&\Pi\left(p^2\right)\left(-g_{\mu \nu}+\frac{p_\mu p_\nu}{p^2}\right)+\cdots\, ,
\end{eqnarray}
where the pole residue $\lambda_Z$ is defined by $\langle0|J_\mu(0)|Z(p)\rangle=\lambda_Z \varepsilon_\mu$,
the $\varepsilon_\mu$ is the polarization vector of the vector hexaquark state $Z$.

At the quark level, we calculate the correlation function $\Pi_{\mu\nu}\left(p\right)$
via the operator product expansion method in perturbative QCD.
The $u$, $d$, $s$ and $c$ quark fields are contracted with the Wick theorem,
and the following result is obtained:
\begin{eqnarray}
&&\Pi_{\mu \nu}\left(p\right)=-i\int d^{4}x e^{ip\cdot x}\nonumber\\
&&\left\{
{\rm Tr}\left[\gamma_\mu D^{nn'}(x)\gamma_\nu C^{n'n}(-x)\right]
{\rm Tr}\left[i\gamma_5 C^{mm'}(x)i\gamma_5U^{m'm}(-x)\right]
{\rm Tr}\left[i\gamma_5S^{kk'}(x)i\gamma_5U^{k'k}(-x)\right]\right. \nonumber\\
&&\left.-{\rm Tr}\left[\gamma_\mu D^{nn'}(x)\gamma_\nu C^{n'n}(-x)\right]
{\rm Tr}\left[i\gamma_5 C^{mm'}(x)i\gamma_5U^{m'k}(-x)
i\gamma_5S^{kk'}(x)i\gamma_5U^{k'm}(-x)\right]
\right\}\,,
\end{eqnarray}
where the $U_{i j}(x)$, $D_{i j}(x)$, $S_{i j}(x)$ and $C_{i j}(x)$ are the full $u$, $d$, $s$ and $c$ quark propagators,  respectively.
We give the full quark propagators explicitly in the following,
(the $P_{i j}(x)$ denotes the $U_{i j}(x)$ or $D_{i j}(x)$ ),
\begin{eqnarray}
P_{i j}(x)&=&\frac{i\delta_{i j}x\!\!\!/}{2\pi^{2}x^{4}}-\frac{\delta_{i j}\langle\bar{q}q\rangle}{12}-\frac{\delta_{i j}x^{2}\langle\bar{q}g_{s}\sigma Gq\rangle}{192}-\frac{i g_{s}G_{\alpha\beta}^{n}t_{i j}^{n}(x\!\!\!/\sigma^{\alpha\beta}+\sigma^{\alpha\beta}x\!\!\!/)}{32\pi^{2}x^{2}} \nonumber\\
&&-\frac{1}{8}\langle\bar{q}_{j}\sigma^{\alpha\beta}q_{i}\rangle\sigma_{\alpha\beta}
+\cdots \ ,\label{propagator-u,d}\\
S_{i j}(x)&=&\frac{i\delta_{i j}x\!\!\!/}{2\pi^{2}x^{4}}-\frac{\delta_{i j}m_{s}}{4\pi^{2}x^{2}}-\frac{\delta_{i j}\langle\bar{s}s\rangle}{12}+\frac{i\delta_{i j}x\!\!\!/m_{s}\langle\bar{s}s\rangle}{48}-\frac{\delta_{i j}x^{2}\langle\bar{s}g_{s}\sigma Gs\rangle}{192} \nonumber\\
&&+\frac{i\delta_{i j}x^{2}x\!\!\!/m_{s}\langle\bar{s}g_{s}\sigma Gs\rangle}{1152}
-\frac{i g_{s}G_{\alpha\beta}^{n}t_{i j}^{n}(x\!\!\!/\sigma^{\alpha\beta}+\sigma^{\alpha\beta}x\!\!\!/)}{32\pi^{2}x^{2}} \nonumber\\
&&-\frac{1}{8}\langle\bar{s}_{j}\sigma^{\alpha\beta}s_{i}\rangle\sigma_{\alpha\beta}
+\cdots \ ,\label{propagator-s}\\
C_{i j}(x)&=&\frac{i}{(2\pi)^4}\int d^4 ke^{-ik\cdot x}\bigg\{\frac{k\!\!\!/ +m_{c}}{k^{2}-m_{c}^{2}}\delta_{i j}-g_{s}t_{i j}^{n}G_{\alpha\beta}^{n}\frac{(k\!\!\!/+m_{c})\sigma^{\alpha\beta}+\sigma^{\alpha\beta}(k\!\!\!/+m_{c})}{4(k^{2}-m_{c}^{2})^{2}} \nonumber\\
&&-\frac{g_{s}^{2}(t^{n}t^{m})_{i j}G_{\alpha\beta}^{n}G_{\mu\nu}^{n}(f^{\alpha\beta\mu\nu}+f^{\alpha\mu\beta\nu}+f^{\alpha\mu\nu\beta})}{4(k^{2}-m_{c}^2)^{5}}+\cdots\bigg\} \ ,
\end{eqnarray}
\begin{eqnarray}
f^{\lambda\alpha\beta}&=&(k\!\!\!/+m_{c})\gamma^{\lambda}(k\!\!\!/+m_{c})\gamma^{\alpha}(k\!\!\!/+m_{c})\gamma^{\beta}(k\!\!\!/+m_{c})\ ,\nonumber\\
f^{\alpha\beta\mu\nu}&=&(k\!\!\!/+m_{c})\gamma^{\alpha}(k\!\!\!/+m_{c})\gamma^{\beta}(k\!\!\!/+m_{c})\gamma^{\mu}(k\!\!\!/+m_{c})\gamma^{\nu}(k\!\!\!/+m_{c})\ ,
\end{eqnarray}
and $t^{n}=\frac{\lambda^{n}}{2}$, the $\lambda^{n}$ is the
Gell-Mann matrix \cite{hadron-side-2}.
We compute the integrals in the coordinate space for the light quark propagators
and in the momentum space for the charm quark propagators,
and obtain the QCD spectral density $\rho(s)$ via taking the imaginary part of the correlation
function: $\rho(s)={\rm lim}_{\varepsilon \to 0}\frac{\text{Im}\Pi(s+i\varepsilon)}{\pi}$
 \cite{axialvector}. In the operator product expansion,
we take into account the contributions of vacuum condensates up to dimension-16, and keep the terms which are linear in the strange quark mass $m_s$.
We take the truncation $k\leq1$ for the operators of the order  $\mathcal{O}( \alpha_s^{k})$ in a consistent way and discard the perturbative corrections.
Furthermore, the condensates $\langle\bar{q}q\rangle \langle\frac{\alpha_s GG}{\pi}\rangle$,
$\langle\bar{q}q\rangle^2 \langle\frac{\alpha_s GG}{\pi}\rangle$
and $\langle\bar{q}q\rangle^3 \langle\frac{\alpha_s GG}{\pi}\rangle$
  play a minor important role and are neglected.

According to the quark-hadron duality, we match the correlation function $\Pi(p^2)$ gotten on the hadron side and at the quark level below the continuum threshold $s_0$, and perform Borel transform with respect to the variable $P^2 =-p^2$
to obtain the   QCD sum rule:
\begin{eqnarray}\label{PoleResidue}
\lambda_Z^{2}\exp\left(-\frac{M_Z^{2}}{T^{2}}\right)
&=&\int_{4m_{c}^{2}}^{s_{0}}ds\rho\left(s\right)\exp\left(-\frac{s}{T^{2}}\right)\ ,
\end{eqnarray}
where the QCD spectral density
\begin{eqnarray}
\rho\left(s\right)&=& \rho_{0}\left(s\right)+\rho_{3}\left(s\right)+\rho_{4}\left(s\right)
+\rho_{5}\left(s\right)+\rho_{6}\left(s\right)+\rho_{8}\left(s\right)
+\rho_{9}\left(s\right)+\rho_{10}\left(s\right)+\rho_{11}\left(s\right) \nonumber\\
&&+\rho_{12}\left(s\right)+\rho_{13}\left(s\right)+\rho_{14}\left(s\right)+\rho_{16}\left(s\right)\ ,
\end{eqnarray}
the subscripts 0, 3, 4, 5, 6, 8, 9, 10, 11, 12, 13, 14, 16 denote the dimensions of the vacuum condensates,
the $T^2$ is the Borel parameter,
the lengthy and complicated expressions are neglected for simplicity.
However,
for the explicit expressions of the QCD special densities,
the interested readers can obtain them through emailing us.

We derive Eq.\eqref{PoleResidue} with respect to $\frac{1}{T^2}$,
and eliminate the pole residue $\lambda_Z$
to extract the QCD sum rule for the mass:
\begin{eqnarray}\label{mass}
M_Z^{2}&=&\frac{\int_{4m_{c}^{2}}^{s_{0}}ds\frac{d}{d\left(-1/T^{2}\right)}\rho\left(s\right)\exp\left(-\frac{s}{T^{2}}\right)} {\int_{4m_{c}^{2}}^{s_{0}}ds\rho\left(s\right)\exp\left(-\frac{s}{T^{2}}\right)}\ .
\end{eqnarray}

\section{Numerical results and discussions}
In this section, we perform the numerical analysis.
To extract the numerical values of $M_Z$,
we take the values of the vacuum condensates
$\langle\bar{q}q\rangle=-(0.24\pm0.01\,\text{GeV})^{3}$, $\langle\bar{s}s\rangle=(0.8\pm0.1)\langle\bar{q}q\rangle$,
$\langle\bar{q}g_{s}\sigma Gq\rangle=m_{0}^{2}\langle\bar{q}q\rangle$,
$\langle\bar{s}g_{s}\sigma Gs\rangle=m_{0}^{2}\langle\bar{s}s\rangle$,
$m_{0}^{2}=(0.8\pm0.1)\,\text{GeV}^{2}$, $\langle\frac{\alpha_{s}GG}{\pi}\rangle=(0.33\,\text{GeV})^{4}$
at the energy scale  $\mu=1\,\text{GeV}$ \cite{hadron-side-1,hadron-side-2,ColangeloReview},
choose the $\overline{MS}$ masses $m_{c}(m_c)=(1.275\pm0.025)\,\rm{GeV}$,
$m_{s}(\mu=2\,\rm{GeV})=(0.095^{+0.009}_{-0.003})\,\rm{GeV}$ from the Particle Data Group \cite{XYZ},
and neglect the up and down quark masses, i.e., $m_u=m_d=0$.
Moreover, we consider the energy-scale dependence of the input parameters
on the QCD side from the renormalization group equation,
\begin{eqnarray}
\langle\bar{q}q\rangle(\mu)&=&\langle\bar{q}q\rangle(1\rm{GeV})\left[\frac{\alpha_{s}(1\rm{GeV})}{\alpha_{s}(\mu)}\right]^{\frac{4}{9}}\, ,\nonumber\\
\langle\bar{s}s\rangle(\mu)&=&\langle\bar{s}s\rangle(1\rm{GeV})\left[\frac{\alpha_{s}(1\rm{GeV})}{\alpha_{s}(\mu)}\right]^{\frac{4}{9}}\, ,\nonumber\\
\langle\bar{q}g_{s}\sigma Gq\rangle(\mu)&=&\langle\bar{q}g_{s}\sigma
Gq\rangle(1\rm{GeV})\left[\frac{\alpha_{s}(1\rm{GeV})}{\alpha_{s}(\mu)}\right]^{\frac{2}{27}}\, ,\nonumber\\
\langle\bar{s}g_{s}\sigma
Gs\rangle(\mu)&=&\langle\bar{s}g_{s}\sigma
Gs\rangle(1\rm{GeV})\left[\frac{\alpha_{s}(1\rm{GeV})}{\alpha_{s}(\mu)}\right]^{\frac{2}{27}}\, ,\nonumber\\
m_{s}(\mu)&=&m_{s}\left(2\,\text{GeV}\right)\left[\frac{\alpha_{s}(\mu)}{\alpha_{s}(2\,\text{GeV})}\right]^{\frac{4}{9}}\, ,\nonumber\\
m_{c}(\mu)&=&m_{c}\left(m_{c}\right)\left[\frac{\alpha_{s}(\mu)}{\alpha_{s}(m_{c})}\right]^{\frac{12}{25}}\, ,\nonumber\\
\alpha_{s}(\mu)&=&\frac{1}{b_{0}t}\left[1-\frac{b_{1}}{b_{0}^{2}}\frac{\log t}{t}+\frac{b_{1}^{2}\left(\log^{2}t-\log t-1\right)+b_{0}b_{2}}{b_{0}^{4}t^{2}}\right]\, ,
\end{eqnarray}
where $t=\log \frac{\mu^{2}}{\Lambda^{2}}$, $b_{0}=\frac{33-2n_{f}}{12\pi}$, $b_{1}=\frac{153-19n_{f}}{24\pi^{2}}$, $b_{2}=\frac{2857-\frac{5033}{9}n_{f}+\frac{325}{27}n_{f}^{2}}{128\pi^{3}}$,
$\Lambda=213\,\text{MeV}$, $296\,\text{MeV}$ and $339\,\text{MeV}$
for the flavors $n_{f}=5$, $4$ and $3$, respectively \cite{XYZ}.

For the hadron mass, it is independent of the energy scale because of its observability.
However, in calculations, the perturbative corrections are neglected, the operators of the orders $\mathcal{O}_n( \alpha_s^{k})$
with $k>1$ or the dimensions $n>16$ are discarded,
and some higher dimensional vacuum condensates
are factorized into lower dimensional ones therefore the corresponding energy-scale dependence is modified.
We have to take into account the energy-scale dependence of the QCD sum rules.

In Refs.\cite{axialvector,energy-scale-1,energy-scale-2,energy-scale-3},
the energy-scale dependence of the QCD sum rules is studied in detail
for the hidden-charm tetraquark states and molecular states,
and an energy scale formula $\mu=\sqrt{M_{X/Y/Z}^{2}-\left(2{\mathbb{M}}_{c}\right)^{2}}$ is come up with
to determine the optimal energy scale.
This energy-scale formula enhances the pole contribution remarkably, improves the convergent behaviors in the operator product expansion,
and works well for the exotic hadron states.
In this article, we explore the $D\bar{D}^*K$ state $Z$
through constructing the color singlet-singlet-singlet type current based on the color-singlet $q\bar{q}$ substructure.
For the two-meson molecular states, the basic constituent is also the color-singlet $q\bar{q}$ substructure \cite{energy-scale-3}.
Hence, the previous works can be extended to study the $D\bar{D}^*K$ state.
We employ the energy-scale formula $\mu=\sqrt{M_{X/Y/Z}^{2}-\left(2{\mathbb{M}}_{c}\right)^{2}}$
with the updated value of the effective $c$-quark mass ${\mathbb{M}}_{c}=1.85\,\rm{GeV}$
to take the ideal energy scale of the QCD spectral density.

\begin{figure}[htp]
\centering
\includegraphics[totalheight=8cm,width=10cm]{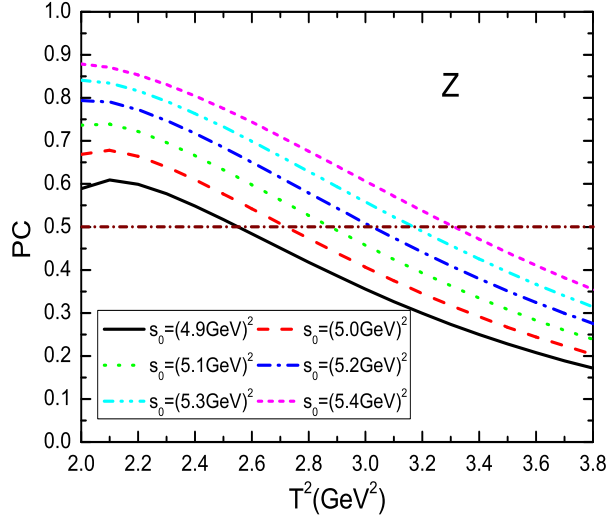}
\caption{The pole contribution with variation of the Borel parameter $T^{2}$.}\label{fig:fig1}
\end{figure}

At the present time, no candidate is observed  experimentally for the hexaquark state $Z$ with the symbolic quark constituent $c\bar{c}d\bar{u}s\bar{u}$.
However, in the scenario of four-quark states,
the $Z_c(3900)$ and $Z(4430)$ can be tentatively assigned to be the ground state
and the first radial excited state of the axialvector four-quark states, respectively \cite{CTP325},
while the $X(3915)$ and $X(4500)$ can be tentatively assigned to be the ground state
and the first radial excited state of the scalar four-quark states, respectively \cite{EPJ-C77-A53}.
By comparison,
the energy gap is about $0.6\, \text{GeV}$ between the ground state and the first radial excited state
of the hidden-charm four-quark states.
Here,
we suppose the energy gap is also about $0.6\, \text{GeV}$ between the ground state and the first radial excited state
of the hidden-charm six-quark states,
and take the relation $\sqrt{s_{0}}=M_Z+(0.4-0.6)\,\text{GeV}$
as a constraint to obey.

\begin{figure}[htp]
\centering
\includegraphics[totalheight=8cm,width=10cm]{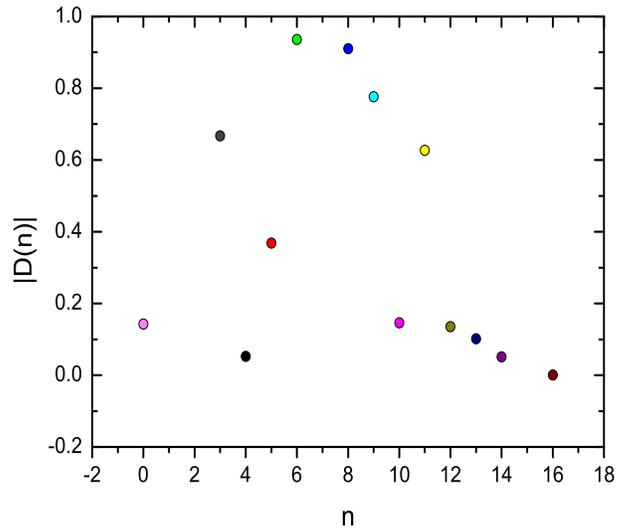}
\caption{The absolute contributions of the vacuum condensates with dimension $n$ in the operator product expansion.}\label{fig:fig2}
\end{figure}

In Eq.\eqref{mass}, there are two free parameters: the Borel parameter $T^2$
and the continuum threshold parameter $s_0$.
The extracted hadron mass is a function of the Borel parameter $T^2$
and the continuum threshold parameter $s_0$.
To obtain a reliable mass sum rule analysis,
we obey two criteria
to choose suitable working ranges for the two free parameters.
One criterion is the pole dominance on the phenomenological side,
which requires the pole contribution (PC) to be about $(40-60)\%$.
The PC is defined as:
\begin{eqnarray}
\text{PC}&=&\frac{\int_{4m_{c}^{2}}^{s_{0}}ds\rho\left(s\right)\exp\left(-\frac{s}{T^{2}}\right)} {\int_{4m_{c}^{2}}^{\infty}ds\rho\left(s\right)\exp\left(-\frac{s}{T^{2}}\right)}\ .
\end{eqnarray}
The other criterion is the convergence of the operator product expansion. To judge the
convergence, we compute the contributions of the vacuum condensates $D(n)$
in the operator product expansion with the formula:
\begin{eqnarray}
D(n)&=&\frac{\int_{4m_{c}^{2}}^{s_{0}}ds\rho_{n}(s)\exp\left(-\frac{s}{T^{2}}\right)}
{\int_{4m_{c}^{2}}^{s_{0}}ds\rho\left(s\right)\exp\left(-\frac{s}{T^{2}}\right)}\ ,
\end{eqnarray}
where the $n$ is the dimension of the vacuum condensates.

\begin{figure}[htp]
\centering
\includegraphics[totalheight=8cm,width=10cm]{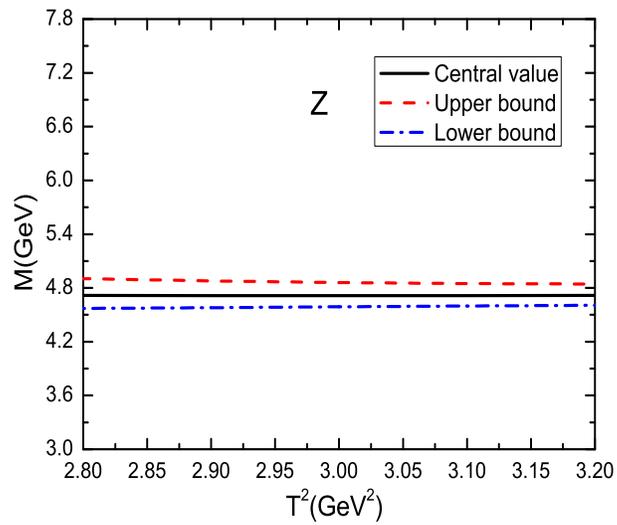}
\caption{The mass with variation of the Borel parameter $T^{2}$.}\label{fig:fig3}
\end{figure}

\begin{figure}[htp]
\centering
\includegraphics[totalheight=8cm,width=10cm]{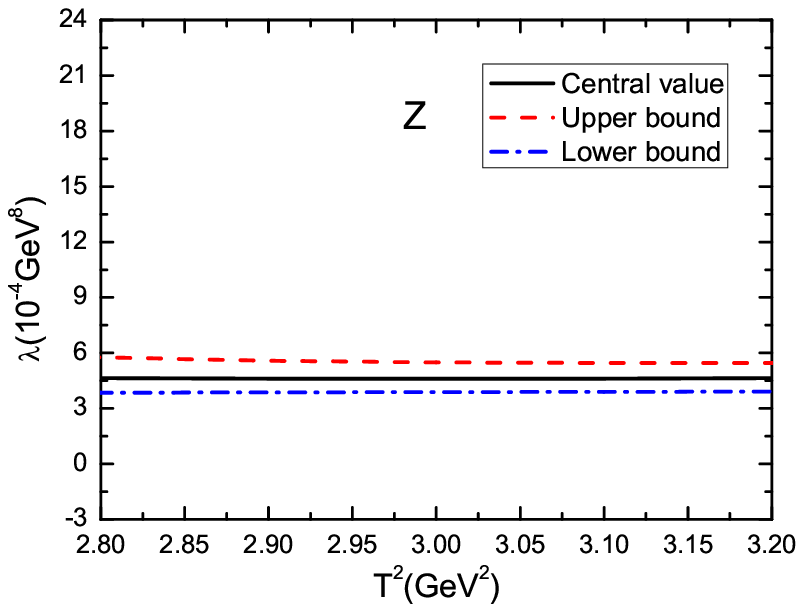}
\caption{The pole residue with variation of the Borel parameters $T^{2}$.}\label{fig:fig4}
\end{figure}

In Fig. \ref{fig:fig1},
we show the variation of the PC
with respect to the Borel parameter $T^2$
for different values of the continuum threshold parameter $s_0$
at the energy scale $\mu=2.9\,\text{GeV}$.
From the figure,
we can see that the value $\sqrt{s_0}\leq5.0\,\rm{GeV}$
is too tiny to obey the pole dominance criterion
and result in sound Borel window for the state $Z$.
To warrant the Borel platform for the mass $m_Z$,
we take the value $T^2=(2.8-3.2)\,\rm{GeV}^2$.
In the above Borel window, if we choose the value $\sqrt{s_0}=(5.1-5.3)\,\rm{GeV}$,
the PC is about $(39-63)\%$.
The pole dominance condition is well satisfied.

In Fig.\ref{fig:fig2},
we draw the absolute contribution values of the vacuum condensates $|D(n)|$
at central values of the above input parameters.
From the figure, we can observe that
the contribution of the perturbative term $D(0)$ is not the dominant contribution,
the contributions of the vacuum condensates with dimensions 3, 6, 8, 9 and 11 are very great.
If we take the contribution of the vacuum condensate with dimension 11 as a milestone,
the absolute contribution values of the vacuum condensates $|D(n)|$ decrease quickly  with the increase of the dimensions $n$,
the operator product expansion converges nicely.

Thus,
we obtain the values $T^2=(2.8-3.2)\,\rm{GeV}^2$,
$\sqrt{s_0}=(5.1-5.3)\,\rm{GeV}$ and $\mu=2.9\,\text{GeV}$ for the state $Z$.
Considering all uncertainties of the input parameters,
we get the values of the mass and pole residue of the state $Z$:
\begin{eqnarray}
M_Z&=&4.71_{-0.11}^{+0.19}\,\rm{GeV}\ ,\nonumber\\
\lambda_Z&=&\left(4.60_{-0.69}^{+1.15}\right)\times 10^{-4}\,\rm{GeV}^8\ ,
\end{eqnarray}
which are shown explicitly in Figs.\ref{fig:fig3}--\ref{fig:fig4}.
Obviously, the energy-scale formula $\mu =\sqrt{M_{X/Y/Z}^2-({2\mathbb{M}}_c)^2}$
and the relation $\sqrt{s_{0}}=M_Z+(0.4-0.6)\,\text{GeV}$ are also well satisfied.
The central value $M_Z=4.71\,\rm{GeV}$ is about $337\,\rm{MeV}$ above the threshold
$M_{K+D+\bar{D}^*}=497.6+1865+2010=4372.6\,\rm{MeV}$,
which indicates that
the $Z$ is probably a resonance state.
For some exotic resonances,
the authors have combined the effective range expansion, unitarity,
analyticity and compositeness coefficient to probe their inner structure
in Refs.\cite{ERE-1,ERE-2}.
Their studies indicated that
the underlying two-particle component (in the present case, corresponding to three-particle component)
play an important or  minor role,
in other words, there are the other hadronic degrees of freedom
inside the corresponding resonance.
Hence, a resonance state embodies the net effect.
Considering the conservation of the angular momentum, parity and isospin,
we list out the possible hadronic decay patterns of the hexaquark state $Z$:
\begin{eqnarray}
Z\longrightarrow J/\psi \pi K,  \eta_c\rho(770)K, D\bar{D}^*K.
\end{eqnarray}
To search for the X(3872), Belle, BaBar and LHCb have collected numerous data in the decay $B\longrightarrow J/\psi \pi \pi K$.
Thus,
the hexaquark state $Z$ may be found by focusing on the most easy channel $J/\psi \pi K$ in the experiment.

\section{Conclusion}
In this article, we construct the color singlet-singlet-singlet interpolating current operator with
$I\left(J^P\right)=\frac{3}{2}\left(1^-\right)$
to study the $D\bar{D}^*K$ system
through QCD sum rules approach by taking into account the contributions of the vacuum condensates up to dimension-16
in the operator product expansion. In numerical calculations, we saturate the hadron side of the QCD sum rule with a hexaquark molecular state,
employ the energy-scale formula
$\mu=\sqrt{M_{X/Y/Z}^{2}-\left(2{\mathbb{M}}_{c}\right)^{2}}$ to take the optimal energy scale of the QCD spectral density,
and seek the ideal  Borel parameter $T^2$ and continuum threshold $s_0$
by obeying two criteria of QCD sum rules for multi-quark states.
Finally, we obtain the mass and pole residue of the corresponding hexaquark molecular state $Z$.
The predicted mass,
$M_Z=4.71_{-0.11}^{+0.19}\,\rm{GeV}$,
which lies above the $D\bar{D}^*K$ threshold,
indicates that the $Z$
is probably a resonance state.
This resonance state $Z$ may be found by focusing on the channel $J/\psi \pi K$
of the decay $B\longrightarrow J/\psi \pi \pi K$ in the future.

\section*{Acknowledgements}
This work is supported by National Natural Science Foundation, Grant Number 11775079.


\begin{thebibliography}{99}

\bibitem{X3872} S. K. Choi, et al., Phys. Rev. Lett. {\bf 91} (2003) 262001.

\bibitem{XYZ} K. A. Olive, et al., Chin. Phys. {\bf C38} (2014) 090001.


\bibitem{RMP90}
S. L. Olsen, T. Skwarnicki and D. Zieminska, Rev. Mod. Phys. {\bf 90} (2018) 15003.

\bibitem{EPJC1534}
N. Brambilla, et al., Eur. Phys. J. {\bf C71} (2011) 1534.


\bibitem{Quark-Model} S. Godfrey and N. Isgur, Phys. Rev. {\bf D32} (1985) 189.

\bibitem{PR639}
H. X. Chen, W. Chen, X. Liu and S. L. Zhu, Phys. Rept. {\bf 639} (2016) 1.

\bibitem{PLB753}
J. He, Phys. Lett. {\bf B753} (2016) 547.

\bibitem{GuoRMP} F. K. Guo, C. Hanhart, U. G. Meissner, Q. Wang, Q. Zhao and B. S. Zou, Rev. Mod. Phys. {\bf 90} (2018) 015004.

\bibitem{PRC69} C. Y. Wong, Phys. Rev. {\bf C69} (2004) 055202.

\bibitem{group}
E. S. Swanson, Phys. Lett. {\bf B588} (2004) 189;
M. Suzuki, Phys. Rev. {\bf D72} (2005) 114013;
M. T. AlFiky, F. Gabbiani and A. A. Petrov, Phys. Lett. {\bf B640} (2006) 238;
S. Fleming, M. Kusunoki, T. Mehen and U. van Kolck, Phys. Rev. {\bf D76} (2007) 034006;
E. Braaten, M. Lu and J. Lee, Phys. Rev. {\bf D76} (2007) 054010;
C. Hanhart, Y. S. Kalashnikova, A. E. Kudryavtsev and A. V. Nefediev, Phys. Rev. {\bf D76} (2007) 034007;
M. B. Voloshin, Phys. Rev. {\bf D76} (2007) 014007;
P. Colangelo, F. De Fazio and S. Nicotri, Phys. Lett. {\bf B650} (2007) 166.


\bibitem{KDD-1} L. Ma, Q. Wang and U.-G. Meissner, arXiv:1711.06143.

\bibitem{KDD-2} X. L. Ren, B. B. Malabarba, L. S. Geng, K. P. Khemchandani  and A. M. Torres, Phys. Lett. {\bf B785} (2018) 112.

\bibitem{QCD-approach}
R. M. Albuquerque, et al., arXiv:1812.08207.

\bibitem{PRD83}
W. Chen and S. L. Zhu, Phys. Rev. {\bf D83} (2011) 034010;
W. Chen, T. G. Steele, H. X. Chen and S. L. Zhu, Eur. Phys. J. {\bf C75} (2015) 358.

\bibitem{CTP54}
J. R. Zhang and M. Q. Huang, Commun. Theor. Phys. {\bf 54} (2010) 1075.

\bibitem{CTP191}
Z. Y. Di, Z. G. Wang, J. X. Zhang and G. L. Yu, Commun. Theor. Phys. {\bf 69} (2018) 191;
Z. G. Wang, Mod. Phys. Lett. {\bf A29} (2014) 1450207.


\bibitem{axialvector} Z. G. Wang and T. Huang, Phys. Rev. {\bf D89} (2014) 054019.

\bibitem{WangZhang-Solid} Z. G. Wang and  J. X. Zhang, Eur. Phys. J. {\bf C78} (2018) 14.



\bibitem{hadron-side-1}
M. A. Shifman, A. I. Vainshtein and V. I. Zakharov, Nucl. Phys. {\bf B147} (1979) 385, 448.

\bibitem{hadron-side-2}
L. J. Reinders, H. Rubinstein and S. Yazaki, Phys. Rept. {\bf 127} (1985) 1.



\bibitem{ColangeloReview}
P. Colangelo and A. Khodjamirian, At the Frontier of Particle Physics: Handbook of QCD,
Vol. {\bf 3}, ed. M. Shifman (World Scientific, Singapore, 2001), p. 1495.


\bibitem{energy-scale-1}
Z. G. Wang, Eur. Phys. J. {\bf C74} (2014) 2874.


\bibitem{energy-scale-2}
Z. G. Wang and T. Huang, Nucl. Phys. {\bf A930} (2014) 63.


\bibitem{energy-scale-3}
Z. G. Wang and T. Huang, Eur. Phys. J. {\bf C74} (2014) 2891;
Z. G. Wang, Eur. Phys. J. {\bf C74} (2014) 2963.


\bibitem{CTP325}
Z. G. Wang, Commun. Theor. Phys. {\bf 63} (2015) 325.

\bibitem{EPJ-C77-A53}
Z. G. Wang, Eur. Phys. J. {\bf C77} (2017) 78;
Z. G. Wang, Eur. Phys. J. {\bf A53} (2017) 19.


\bibitem{ERE-1}
X. W. Kang, Z. H. Guo and J. A. Oller, Phys. Rev. {\bf D94} (2016) 014012.

\bibitem{ERE-2}
R. Gao, Z. H. Guo, X. W. Kang and J. A. Oller, arXiv:1812.07323.



\end{thebibliography}
\end{document}